\documentclass[aip,jcp,amsmath,amssymb,reprint]{revtex4-1}
\usepackage{graphicx}
\usepackage{amssymb}
\usepackage{amsmath}
\usepackage{color}
\usepackage[]{hyperref}
\usepackage{bm}
\usepackage{upgreek}
\usepackage[multidot]{grffile}
\usepackage{natbib}
\usepackage{tikz}
\usepackage{relsize}
\usepackage{subcaption}
\usepackage[utf8]{inputenc}
\usepackage{changes}

\tikzset{fontscale/.style = {font=\relsize{#1}}}

\newcommand{\todo}[1]{{\color{black} #1}}

\begin{document}

\title{Detachment Energies of Spheroidal Particles from Fluid-Fluid Interfaces}
\

\author{Gary B. Davies}
\email{g.davies.11@ucl.ac.uk}
\affiliation{Centre for Computational Science, University College London, 20 Gordon Street, London WC1H 0AJ, United Kingdom.}

\author{Timm Kr\"uger}
\email{timm.krueger@ed.ac.uk}
\affiliation{Institute for Materials and Processes, Department of Engineering, University of Edinburgh, Scotland, Edinburgh EH9 3JL, United Kingdom.}

\author{Peter V. Coveney}
\email{p.v.coveney@ucl.ac.uk}
\affiliation{Centre for Computational Science, University College London, 20 Gordon Street, London WC1H 0AJ, United Kingdom.}

\author{Jens Harting}
\email{j.harting@tue.nl}
\affiliation{Department of Applied Physics, Eindhoven University of Technology, P.O. Box 513, 5600
MB Eindhoven, The Netherlands.}
\affiliation{Faculty of Science and Technology, Mesa+ Institute, University of
Twente, 7500 AE Enschede, The Netherlands.}

\begin{abstract}
The energy required to detach a single particle from a fluid-fluid interface
is an important parameter for designing certain soft materials, for example, emulsions
stabilised by colloidal particles, colloidosomes designed for targeted drug
delivery, and bio-sensors composed of magnetic particles adsorbed at interfaces. For a fixed particle volume, prolate and oblate spheroids attach 
more strongly to interfaces because they have larger particle-interface areas.
Calculating the detachment energy of spheroids necessitates the
difficult measurement of particle-liquid surface tensions, in contrast with
spheres, where the contact angle suffices. We develop a simplified detachment
energy model for spheroids which depends only on the particle aspect ratio and the
height of the particle centre of mass above the fluid-fluid interface. We use lattice Boltzmann simulations to validate the model and provide quantitative evidence that the approach can be applied to simulate particle-stabilized emulsions, and highlight the experimental implications of this validation.  
\end{abstract}

\maketitle

\section{Introduction}

At the turn of the 20th century, Ramsden and Pickering discovered that
colloidal particles stabilise droplets in oil-water
mixtures;~\cite{ramsden_separation_1903,pickering_cxcvi.emulsions_1907} in
2005, researchers predicted the existence of the bicontinuous interfacially
jammed emulsion gel (bijel) using lattice Boltzmann simulations (later
confirmed
experimentally);~\cite{stratford_colloidal_2005,kim_arrest_2008,herzig_bicontinuous_2007}
and in 2011, researchers found that adding tiny amounts of immiscible secondary
fluid to a suspension leads to remarkable changes in its
rheology.~\cite{koos_capillary_2011} These examples show that multiple fluids
interacting with immersed particles can produce complex materials.

Particles adsorb at fluid-fluid interfaces because they lower the free energy,
$F_{\gamma} = \oint_{\partial A} \gamma \,\rm{d}A$, where $\gamma$ is the
surface tension and $\partial A$ the interface area. They do this by replacing
fluid-fluid surface area with particle-fluid surface area, which has a lower
surface tension. Surfactants adsorb at fluid-fluid interfaces because they are amphiphilic. 
However, the free energy reduction due to particle adsorption can be
orders of magnitude larger than the thermal energy, $k_B T$, meaning that
particle adsorption is irreversible. In contrast, soluble surfactant molecules are
usually able to freely adsorb at and desorb from an interface. This means that
particles are often able to stabilise emulsions better, giving rise to
important differences between surfactant-stabilised and particle-stabilised
emulsions.~\cite{binks_particles_2002}

The detachment energy of a single particle from a fluid-fluid interface plays a
vital role in our understanding of particle-stabilised emulsions and, for
example, flotation processes, whereby particles selectively attach to bubbles
depending on their contact angle, isolating the desired mineral. Previous
detachment energy studies focussing on free energy differences between an
equilibrated particle at an interface (buoyancy, gravity and surface-tension
forces interact to determine a particle's equilibrium position at an
interface)~\cite{scheludko_attachment_1976,neumann_spelt_1996,kralchevsky_and_nagayama_2001,
rapacchietta_force_1977,rapacchietta_force_1977-1,singh_fluid_2005,joseph_particle_2003,ivanov_film_1986}
and in the bulk revealed a crucial dependence on particle shape: prolate and
oblate spheroidal particles attach to interfaces more strongly because they
reduce the interface area more than spherical particles for a given particle
volume.~\cite{aveyard_particle_1996,faraudo_stability_2003,koretsky_1971,
levine_stabilization_1989,tadros_vincent_1983,Guzowski2011c}\\\indent
For a
particle already adsorbed at an interface to detach itself, the particle must
deform the interface and overcome the interface's resistive force: there is a
free-energy barrier and an associated activation energy. These energy
contributions are difficult to investigate theoretically. Scheludko et
al.~\cite{scheludko_attachment_1976} and Rapacchietta et
al.~\cite{rapacchietta_force_1977-1} developed analytical expressions
describing the interface deformation for small Bond numbers (ratio of gravity
forces to surface-tension forces). O'Brien ~\cite{obrien_meniscus_1996} showed
that for small Bond numbers the interface's resistive force is linearly
proportional to the particle displacement, similar to the Hooke's law model of
de Gennes et al. ~\cite{de_gennes_wetting:_1985,joanny_model_1984}
Experimentally, Pitois et al.~\cite{pitois_small_2002} measured the detachment
energy of spherical particles from liquid-gas interfaces by integrating
force-displacement curves, a technique we use in this paper, while others were
able to obtain the detachment force but not the detachment
energy.~\cite{preuss_direct_1999,butt_technique_1994} However, none of these
studies extended their treatment to the case of anisotropic spheroidal
particles, which this paper focusses on. 

In this paper, we simulate the detachment of spherical and spheroidal particles from a liquid-liquid interface using a Shan-Chen multicomponent lattice Boltzmann (LB) model.~\cite{shan_lattice_1993,shan_simulation_1994,jansen_bijels_2011,frijters_effects_2012}
LB simulations~\cite{chen_lattice_1998,shan_simulation_1994,shan_lattice_1993,orlandini_lattice_1995,swift_lattice_1996,ladd_lattice-boltzmann_2001} can play an important role in understanding the fundamental interactions 
between particles and interfaces and elucidate the behaviour of macroscopic
systems such as Pickering emulsions, bijels and capillary
suspensions.~\cite{jansen_bijels_2011,frijters_effects_2012,guenther-timescales_2014,kim_bijels_2010,joshi_multiphase_2009,bib:jens-floriang-2013}
In the Shan-Chen multicomponent LB model~\cite{shan_lattice_1993,shan_simulation_1994} that we utilise in this paper, surface-tension emerges from the
fundamental mesoscopic interactions of particle distribution functions that the
algorithm describes. No assumptions are made about the dynamics of the
contact line during detachment. 

We develop a simple thermodynamic model for the detachment energy of spheroidal particles from fluid-fluid interfaces as a function of contact angle and aspect ratio only, and highlight the implications of our simplifications. 

This paper is organised as follows. Section \ref{chap:theory_part} describes previous thermodynamic models for the detachment energy of spherical and spheroidal particles, Section \ref{chap:simmethod} details our simulation methods. The main results are presented in Section \ref{chap:results} and Section \ref{chap:conclusions} concludes the article. 

\section{Thermodynamic Models of Particles Adsorbed at Fluid-Fluid Interfaces}
\label{chap:theory_part}
\subsection{Spherical Particles}

\begin{figure}
\begin{center}
\includegraphics[width=0.45\textwidth]{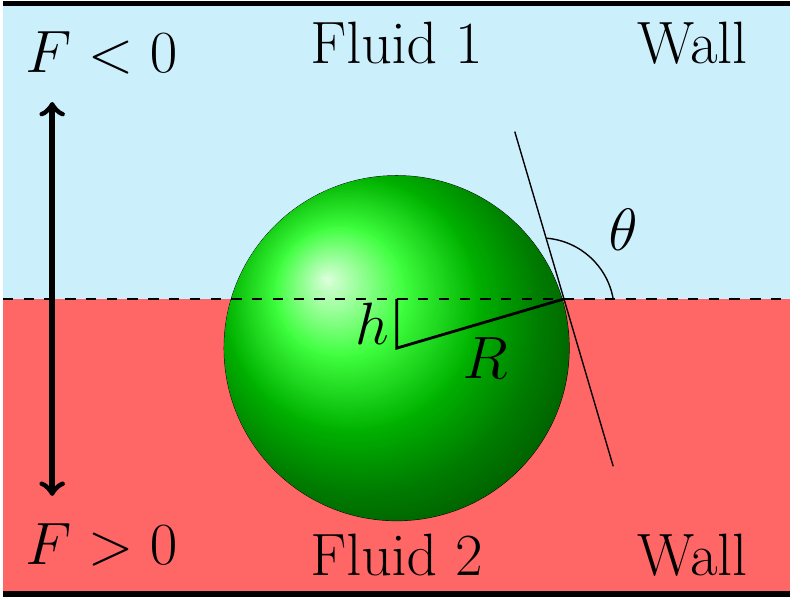}
\caption{\raggedright Equilibrium state of a spherical particle at a fluid-fluid interface. The
contact angle, $\theta = \cos^{-1}(h/R)$, where $h$ is the height of the
particle centre of mass above the interface and $R$ the particle radius, is
defined with respect to Fluid 1. The force, $F$, on the particle acts to
detach it into the wetting (positive force) or non-wetting (negative force)
fluid. }
\label{pic:contact}
\end{center}
\end{figure}

The ratio of gravity forces to surface tension forces for a particle adsorbed at an interface is called the Bond number
\begin{align}
\mathrm{Bo} = \frac{(\rho_{p}-\rho_{f}) g d^2}{\gamma}
\end{align}
\noindent where $\rho_{p}$ and $\rho_{f}$ are the particle and fluid densities respectively, $d$ is the characteristic particle size and $g$ is acceleration due to gravity. For particles of micron size, $\mathrm{Bo} \ll  1$ and surface-tension forces dominate. In this case, the surface free energy of a particle at an interface (Fig.~\ref{pic:contact}) is given by

\begin{align}
\label{anal_interface}
E = \gamma_{12}A_{12} + \gamma_{p1}A_{p1} + \gamma_{p2}A_{p2} 
\end{align}

\noindent where $A_{ij}$ is the area of the $i,j$ interface and $\gamma_{ij}$ is the surface-energy of the $i,j$ interface where $i,j =$ \{1: fluid $1$, 2: fluid $2$, p: particle\}.~\cite{binks_horozov_2006} We neglect line-tension since it is relevant only for nano-sized particles.~\cite{faraudo_stability_2003} The surface area of the particle is $A_{p} = A_{p1} + A_{p2}$. The free energy of a system in which the particle is fully immersed in either fluid $1$ or fluid $2$ is given by $E_{i} = \gamma_{12}A_{12} + \gamma_{pi}A_{pi}$ where $i=\mathrm{1, 2}$.

Taking the free energy difference between a spherical particle at an interface (Fig. \ref{pic:contact}) and a spherical particle immersed in the bulk fluid yields the detachment energy~\cite{koretsky_1971,levine_stabilization_1989,tadros_vincent_1983}

\begin{align}
\label{detachment_energy}
E = \pi R^2 \gamma_{12}(1-| \cos\theta |)^2. 
\end{align}

For neutrally wetting micron-sized particles at an interface with surface-tension $\gamma_{12} = 50\, \mathrm{mN}\, \mathrm{m}^{-1}$, the detachment energy is much larger than the thermal energy, \todo{$E / k_B T \sim 10^7$}, and particles irreversibly attach to the interface. For nano-sized particles at the same interface with very large or small contact angles, $E \sim k_B T$, and particles may freely adsorb at and desorb from the interface, \todo{similar to soluble surfactant molecules.}~\cite{binks_particles_2002}

\subsection{Oblate and Prolate Spheroids}
\label{sec:spheroids}

\begin{figure}
\includegraphics[width=0.45\textwidth]{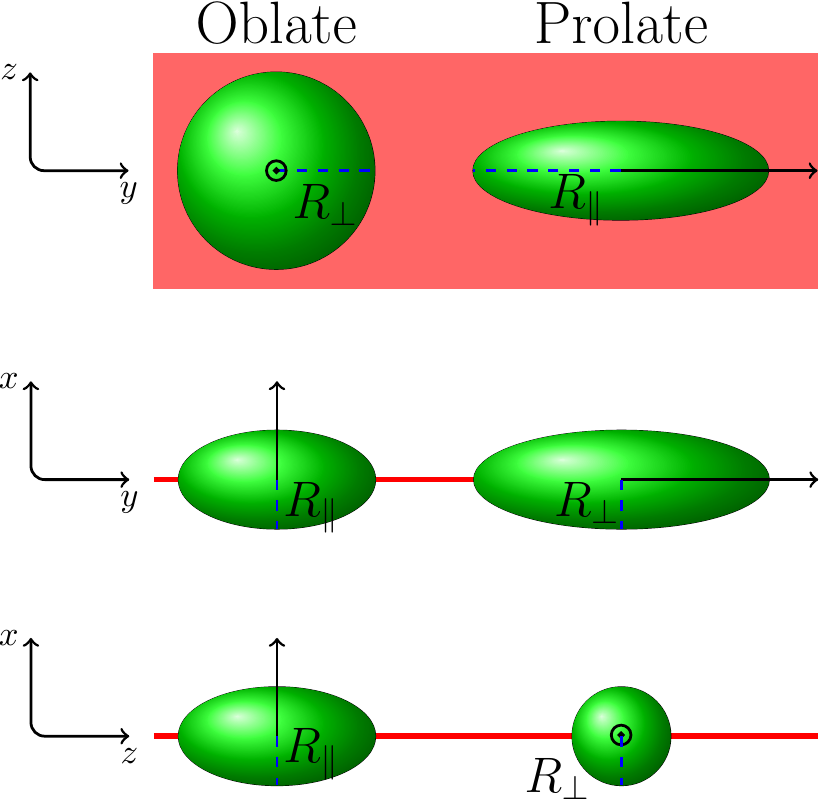}

	\caption{\raggedright Illustration of oblate ($\alpha=0.5$) and prolate ($\alpha=3$) spheroidal particles adsorbed at a
fluid-fluid interface. $R_{\perp}$ and $R_{\|}$ are the radii perpendicular
and parallel to the symmetry axis (black arrows), respectively. Top: Overhead
view of particles in equilibrium at the interface (red). Middle and Bottom: Side view
of particles adsorbed to the interface (red), \todo{showing that the configuration of the particle at the interface is not symmetric for prolate spheroids.}}
	\label{pic:spheroids}
	 
\end{figure}

\citet{faraudo_stability_2003} developed an explicit analytic expression for
the detachment energy of spheroidal particles from a planar fluid-fluid
interface based on free energy differences. Compared with spherical particles,
the detachment energy additionally depends on the orientation of the particle
at the interface and the particle aspect ratio, $\alpha
= R_{\parallel}/ R_{\perp}$, where $R_{\parallel}$ and $R_\perp$ are the radii parallel
and orthogonal to the particle's symmetry axis, respectively. Particles with $\alpha > 1$ are prolate and particles with $\alpha < 1$ are oblate (Fig.~\ref{pic:spheroids}). The model of~\citet{faraudo_stability_2003} assumes a
flat three-phase contact line. For neutrally wetting spheroidal particles in
their equilibrium configuration the equations are exact; for non-neutrally
wetting prolate spheroids and for non-equilibrium \todo{orientations the configuration of the particle at the interface is not symmetric} (Fig.~\ref{pic:spheroids}) and the particle deforms the
interface and three-phase contact line according to Laplace's equation. This
deformation is the cause of long-ranged quadrupolar capillary interactions
between prolate spheroidal particles adsorbed at fluid-fluid interfaces.~\cite{botto_capillary_2012, loudet_capillary_2005}
The equilibrium orientation
of oblate and prolate spheroidal particles is with their symmetry axes parallel and
perpendicular to the interface normal, respectively. The orientation-dependent
detachment energies are~\cite{faraudo_stability_2003}

\begin{align}
\label{eq:fperp}
\Delta E^{\perp} & = \frac{\alpha}{4G}(1-\bar{h}^2) - \frac{\gamma_{2p} - \gamma_{1p}}{\gamma_{12}} \bar{A}^{\perp}_{2p} (\bar{h}), \\
\label{eq:fpar}
\Delta E^{\parallel} & = \frac{1}{4G}(1-\bar{h}^2) - \frac{\gamma_{2p} - \gamma_{1p}}{\gamma_{12}} \bar{A}^{\parallel}_{2p} (\bar{h})
\end{align}

\noindent where $\bar{h} = h/R_{\perp}$ and $\bar{h} = h/R_{\parallel}$ for prolate and oblate spheroidal particles in their equilibrium orientation, respectively, and $G$ is a geometrical aspect factor:

\begin{align}
G = 
\begin{cases}
\frac{1}{2} + \frac{\alpha^2}{4 \epsilon} \log \frac{1+\epsilon}{1-\epsilon} & \text{if } \alpha \leq1 \\
\frac{1}{2} + \frac{1}{2}\frac{\alpha}{\epsilon} \sin^{-1} \epsilon & \text{if } \alpha \geq 1 \\
\end{cases}
\end{align}

\noindent where $\epsilon = \sqrt{1-\alpha^2}$ and $\epsilon = \sqrt{1-\alpha^{-2}}$ is the eccentricity for oblate and prolate spheroids, respectively. \todo{$\bar{A}^{\perp}_{2p}(\bar{h})$ and $\bar{A}^{\parallel}_{2p}(\bar{h})$ represent the fraction of the particle immersed in fluid 2 and were incorrectly defined by Faraudo and Bresme, due to a typo.~\cite{faraudo_stability_2003} The correct equations are}
 
\begin{align}
\label{ap2perp}
\bar{A}^{\perp}_{2p} &= \frac{ \alpha}{\pi G(\alpha)} \int \limits_{0}^{1} \mathrm{d}x  \sqrt{1 - \left(1-\bar{h}^2\right) \left(1-\alpha^{-2}\right) x^{2}} \\\nonumber 
& \quad \times \sqrt{1-\bar{h}^2} \tan^{-1} \left[\frac{1}{\bar{h}} \sqrt{\left(1 - \bar{h}^2\right) \left(1 - x^{2}\right)}\right], \\
\label{ap2par}
\bar{A}^{\parallel}_{2p}  &= \frac{1}{2} - \frac{\alpha}{4G}\bar{h} \sqrt{1 + \frac{\epsilon^2\bar{h}^2}{\alpha^2}} - \frac{\alpha^2}{4G\epsilon} \sinh^{-1} \left(\frac{\epsilon \bar{h}}{\alpha}\right).
\end{align}

\noindent For a neutrally wetting particle, $\bar{h} = 0$ and Equations (\ref{eq:fperp}) and (\ref{eq:fpar}) become $\Delta E^{\perp} = \frac{\alpha}{4G}$ and $\Delta E^{\parallel} = \frac{1}{4G}$ respectively.

\section{Simulation Model and Methods}
\label{chap:simmethod}
We employ the \todo{lattice Boltzmann} (LB) method on a D3Q19
lattice\cite{bib:qian-dhumieres-lallemand} with the Shan-Chen multi-component
model\cite{shan_lattice_1993,aidun10} for the binary liquid part of the system.
Suspended particles are implemented following the pioneering work of Ladd and
Aidun.\cite{ladd_numerical_1994,ladd_lattice-boltzmann_2001,aidun98,jansen_bijels_2011}
The LB method can be considered an alternative to traditional Navier-Stokes
solvers for fluids and due to its local nature is well suited for
implementation on supercomputers.  
While elaborate descriptions of the model implementation have been published
previously,\cite{jansen_bijels_2011,frijters_effects_2012,bib:jens-floriang-2013,guenther-timescales_2014}
we revise some relevant details for the present work in the following.

In the LB algorithm, each fluid component $c$ obeys the dynamical equation
 \begin{align}
\label{lbe}
f_i^c(\mathbf{x} + \mathbf{c}_i \Delta t, t + \Delta t) = f_i^c(\mathbf{x},t) + \Omega_i(\mathbf{x},t)
\end{align}
where  $i = 1,\ldots,19$ so that $f^c_i(\mathbf{x},t)$ represents the particle distribution function in direction $\mathbf{c}_i$ at lattice coordinate $\mathbf{x}$ and time $t$. $\Omega_i(\mathbf{x},t)$ is a generic collision operator: We use the Bhatnagar-Gross-Krook (BGK) operator~\cite{bhatnagar_model_1954,bib:benzi-succi-vergassola}
\begin{align}
\Omega_i = - {\frac{\Delta t}{\tau}} \left(f_i - f^\mathrm{eq}_i\right)
\end{align}
which has the effect of relaxing the system towards a local equilibrium
distribution function $f^\mathrm{eq}_i$ on a time scale given by $\tau$.  The
equilibrium distribution is a second order discretization of the Maxwell-Boltzmann distribution. 
Apart from the common choice $\Delta x = \Delta t = 1$, \textit{i.e.}
the introduction of ``lattice units'', we set $\tau = 1$ which leads to a
numerical kinematic viscosity $\nu = \frac{1}{6}$ in lattice units.


The Shan-Chen model involves a mean-field interaction force between an
arbitrary number of liquid components $c$ at lattice site $\mathbf{x}$ and time
step $t$:\cite{shan_lattice_1993}
\begin{align}
\mathbf{F}^c_\mathrm{SC}(\mathbf{x}, t) = - \Psi^c(\mathbf{x}, t) \sum_{c'} g_{cc'} \sum_{\mathbf{x}'} \Psi^{c'}(\mathbf{x}', t)(\mathbf{x}'-\mathbf{x}),
\end{align}
where $\sum_{\mathbf{x}'}$ runs over all lattice neighbours of site $\mathbf{x}$ and $g_{cc'}$ is a coupling constant representing the interaction strength between the liquid components. In our case, we use two components and do not allow self-interactions ($c' \not= c$). The sign of $g_{cc'}$ determines whether the liquids exhibit attraction (positive sign as in our case) or repulsion (negative sign). $\Psi^c$ is a pseudo-potential of liquid component $c$ which plays the role of an ``effective mass''. It is a function of the density  $\rho^c$ of component $c$ only:
\begin{align}
\label{pseudo}
\Psi^c(\mathbf{x}, t) = \Psi(\rho^c(\mathbf{x}, t)) = 1 - \exp(-\rho^{c}(\mathbf{x}, t)).
\end{align}
Furthermore, we define a local order parameter $\phi(\mathbf{x}, t) = \rho^1(\mathbf{x},t) - \rho^2(\mathbf{x},t)$ which we call the ``colour" of the binary liquid. The interface between the liquids is defined as the surface with $\phi(\mathbf{x}, t) = 0$.

It is possible to control the contact angle of the particles.\cite{jansen_bijels_2011,bib:jens-floriang-2013} All lattice sites in the outer shell of a particle are filled with a virtual binary liquid which itself is not governed by the LB equation but which participates in the computation of the Shan-Chen interaction forces. The densities of those virtual liquids (vl) are
\begin{equation}
\rho^{(1)}_{\mathrm{vl}} = \bar{\rho}^{(1)} + \Delta \rho, \quad \rho^{(2)}_{\mathrm{vl}} = \bar{\rho}^{(2)} - \Delta \rho,
\end{equation}
where $\bar \rho^{(1)}$ and $\bar \rho^{(2)}$ are suitable averages of the component densities on the surrounding liquid lattice sites.\cite{jansen_bijels_2011} The parameter $\Delta \rho$ can be chosen to set the desired wettability behaviour of the particle surface. $\Delta \rho = 0$ recovers a neutrally wetting particle with a contact angle $\theta = 90^\circ$.


To perform numerical simulations of a colloidal particle detaching from a
liquid-liquid interface, we initialise a system volume of size $128^3$ in
lattice units which is half filled with liquid $1$ and half liquid $2$ of equal density,
($\rho^{(1)} = \rho^{(2)} = 0.7$), such that an interface forms at $x=64$, and a
particle density greater than the density of the two fluids ($\rho_p = 2$), which is an arbitrary choice.
The particle is placed at the interface and is not under the influence of any
external forces such as gravity. The liquid-liquid interface is initialised
linearly, spanning just a single lattice site. We first equilibrate the system until the interface diffuses and the
particle establishes its equilibrium position, and hence its contact angle, on the interface.~\cite{jansen_bijels_2011} After equilibration the interface
spans 5-6 lattice sites. This has to be taken into account when e.g. determining
the radius of a droplet used in the calculation of the surface tension from the
Young-Laplace law.~\cite{frijters_effects_2012}

Then, we apply a constant external force to the particle.  If
the simulation volume is entirely periodic, the particle causes the interface
to translate through the simulation domain, hence we place walls with simple
bounce-back boundary conditions~\cite{chen_lattice_1998} parallel to the
interface and normal to the particle detachment direction at $x=0$ and $x=128$ (Fig.~\ref{pic:contact}).

For a particle adsorbed at a fluid-fluid interface, the contact angle of the particle, $\theta$, quantifies its wettability by the different fluids, and is determined by
\begin{equation}
\cos\theta = \frac{\gamma_{2p} - \gamma_{1p}}{\gamma_{12}}. 
\end{equation}
For neutrally wetting particles, $\gamma_{1p} = \gamma_{2p}$ and the contact angle is $\theta = 90^{\circ}$. In our model, particles with contact angles $\theta < 90^{\circ}$ and $\theta > 90^{\circ}$ are preferentially wetted by liquid $1$ and $2$, respectively (Fig.~\ref{pic:contact}). As stated above, we are able to vary the contact angle of the particles.~\cite{jansen_bijels_2011,frijters_effects_2012} We determine the contact angle by subtracting the height of the particle \todo{centre of mass} above the interface (we linearly interpolate the interface position) and dividing by the particle radius, $\cos \theta = \frac{h}{R}$.

\begin{figure}
	\centering
	\begin{subfigure}{0.34\linewidth}
		\centering				
		\includegraphics[width=\textwidth]{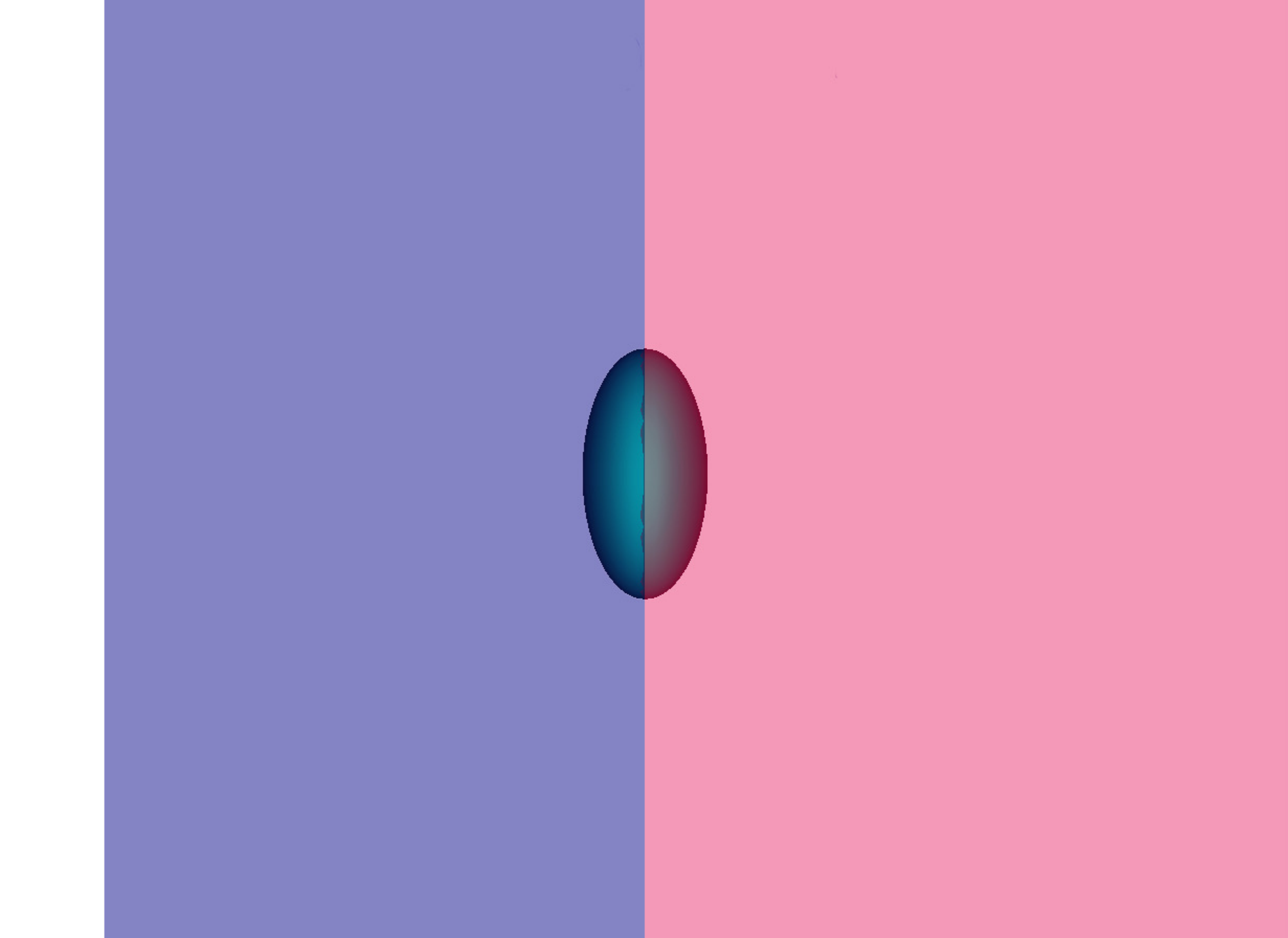}
		\caption{Equilibrium.}
		\label{pic:d0}
	\end{subfigure}
    ~ \hspace{-0.55cm}
	\begin{subfigure}{0.34\linewidth}
		\centering			
		\includegraphics[width=\textwidth]{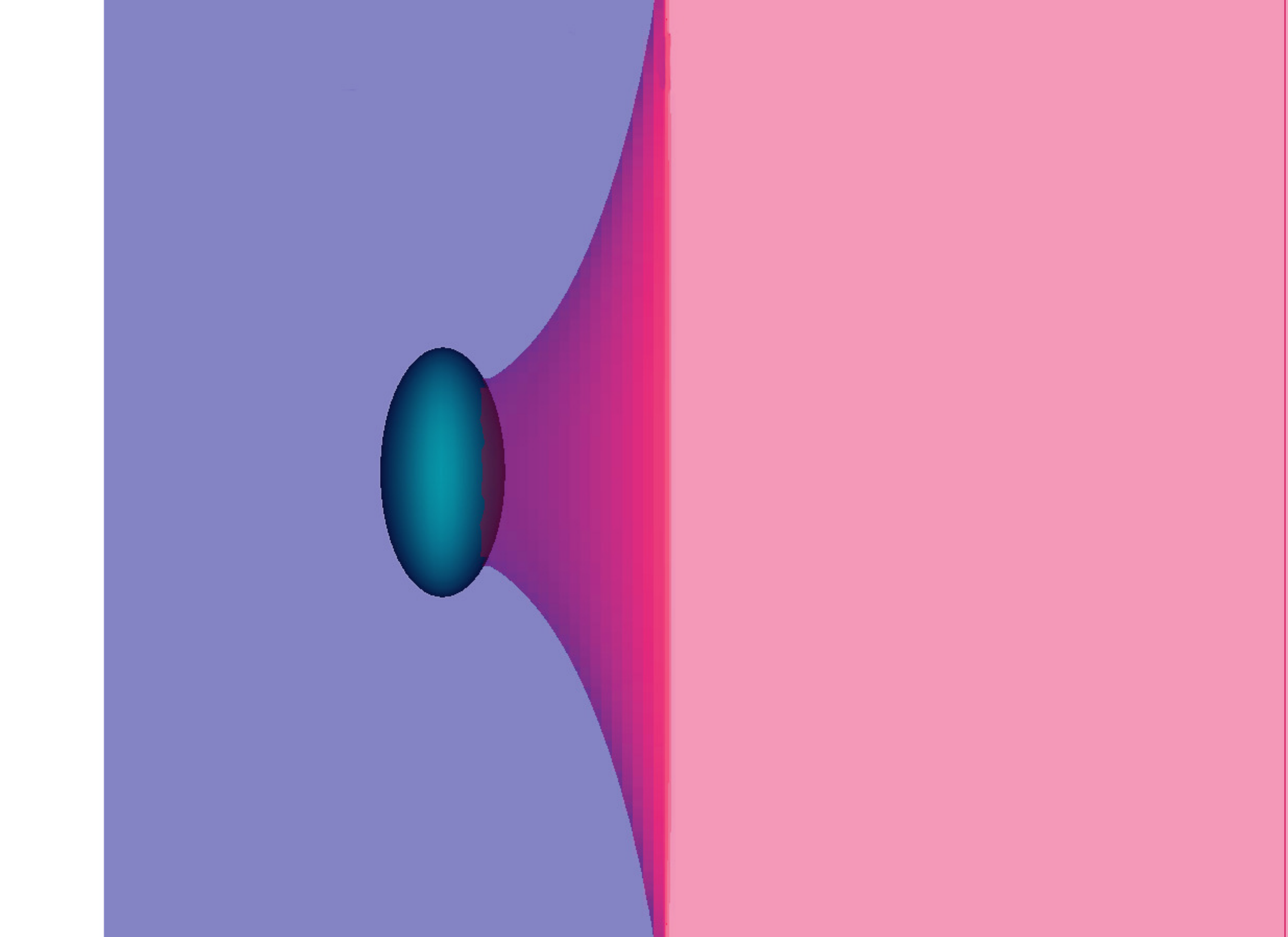} 
		\caption{Resistance.}
		\label{pic:d1}
	\end{subfigure}
    ~ \hspace{-0.55cm}
	\begin{subfigure}{0.34\linewidth}
		\centering				
		\includegraphics[width=\textwidth]{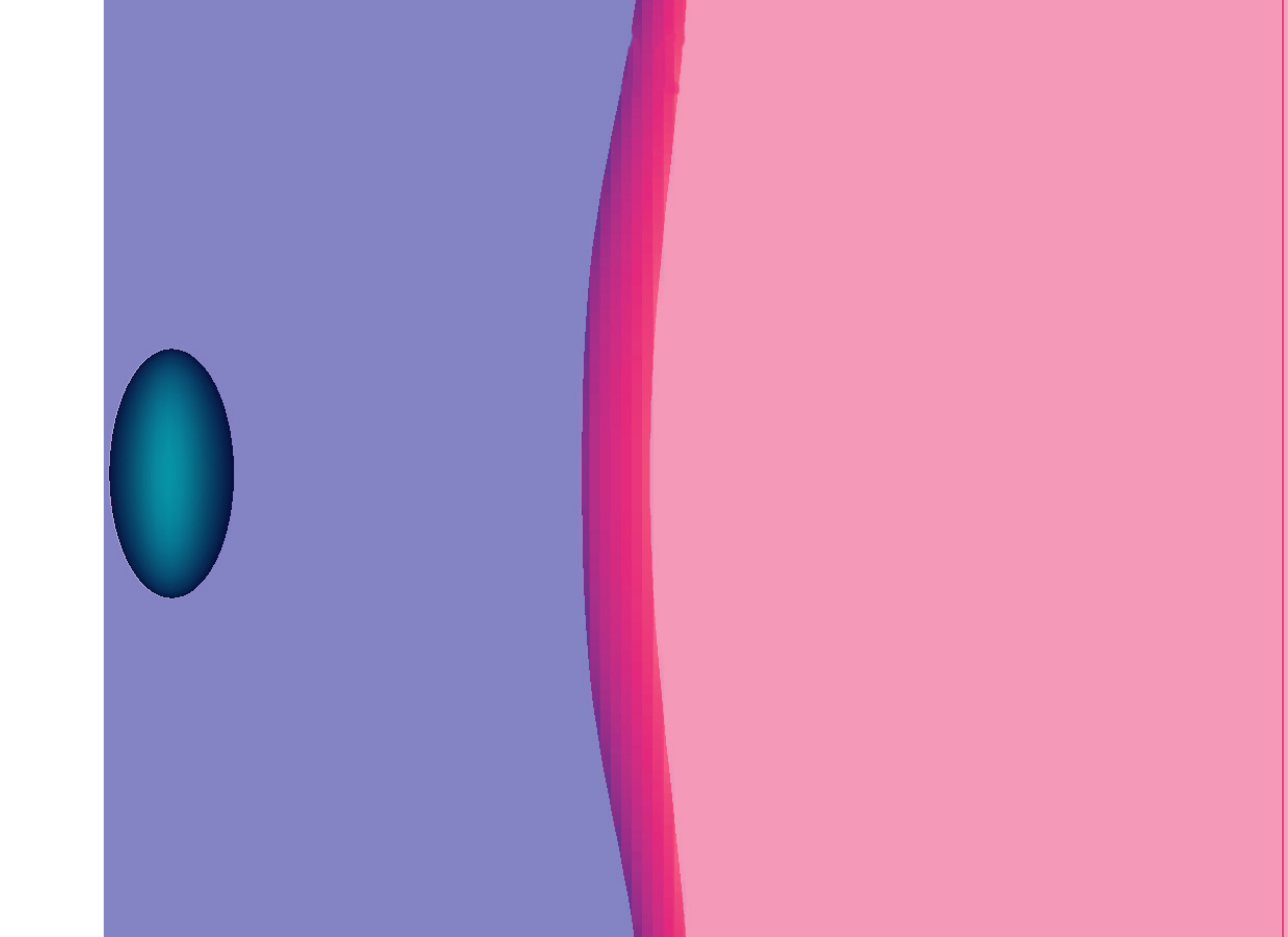}
		\caption{Detachment.}
		\label{pic:d2}
	\end{subfigure}
	\caption{\raggedright Sample snapshots of a prolate spheroidal particle of \todo{aspect ratio} $\alpha=2$ under the influence of an external force detaching from an interface. For each snapshot, we run a simulation with the particle fixed and measure the resistive force from the interface on the particle.}
	\label{pic:oblatedetach}
\end{figure}
In the Shan-Chen multicomponent model~\cite{shan_lattice_1993}, we control the
surface-tension via the fluid-fluid coupling constant, $g_{br}$, which
determines the strength of the interaction between the two fluids. We therefore
need to obtain a mapping from the coupling constant, $g_{br}$, to the surface
tension, $\gamma_{12}$. We use the data from Fig.~2 of Frijters et al.~\cite{frijters_effects_2012} who determined this mapping using the Young-Laplace law. \\\indent
\todo{To see whether a particle detaches for a given applied force, we inspect each simulation manually. To obtain the minimum detachment force, we employ a binary search algorithm: we start the algorithm by using the fact that the particle remains attached at the interface for a zero-force, $F_{\mathrm{att}} = 0$, and guessing a force which detaches the particle, $F_{\mathrm{det}}$. We then run a new simulation with a force, $F_{\mathrm{new}} = \frac{1}{2}(F_{\mathrm{att}} + F_{\mathrm{det}})$ and repeat this procedure until we determine $F_{\mathrm{det}}$ to the desired accuracy. } \\\indent
As a next step, we run a single simulation with the minimum detachment force, saving the simulation state frequently. We then run several simulations from the saved simulation snapshots (Fig.~\ref{pic:oblatedetach}) but now with the particles fixed so that drag, buoyancy and gravity forces can be neglected. We let the systems from each snapshot equilibrate, and we measure the resulting force on the particle which is exactly the resistive force of the interface. This allows us to build a force-distance curve $F(x)$. We fit $F(x)$ with a fourth-order polynomial, which allows us to capture the linear regime and the detachment break-off regime accurately, and integrate the fitted function numerically to obtain the detachment energy. We do this for several particular combinations of surface tension, aspect ratio and contact angle. We find that the integrated detachment energy is insensitive to details of the fitting function, in particular a further increase in the polynomial order. 
The detachment distance is the minimum distance at which the resistive force exerted by the interface on the particle is zero. As discussed shortly, the resistive force decreases discontinuously to zero at the point of detachment. \\\indent
An additional constraint for anisotropic particles is the choice of axis radii since there are an infinite number of axis radii for a given aspect ratio $\alpha = R_{\parallel} / R_{\perp}$.
We vary the particle aspect ratio $\alpha$ while keeping the particle volume, rather than particle surface area, constant.~\cite{binks_horozov_2006} We ensure the minimum axis radius is at least five lattice-sites so that the ratio of the particle diameter to interface thickness is at least $2$:$1$, which has been shown to be sufficient for neutrally wetting spherical particles (Section \ref{chap:results_spherical}), and so that the contact angle is well defined.~\cite{jansen_bijels_2011}

\section{Results}
\label{chap:results}

\subsection{Detachment Energy of Spherical Particles}
\label{chap:results_spherical}
\begin{figure}
\begin{center}
\includegraphics[width=8.6cm]{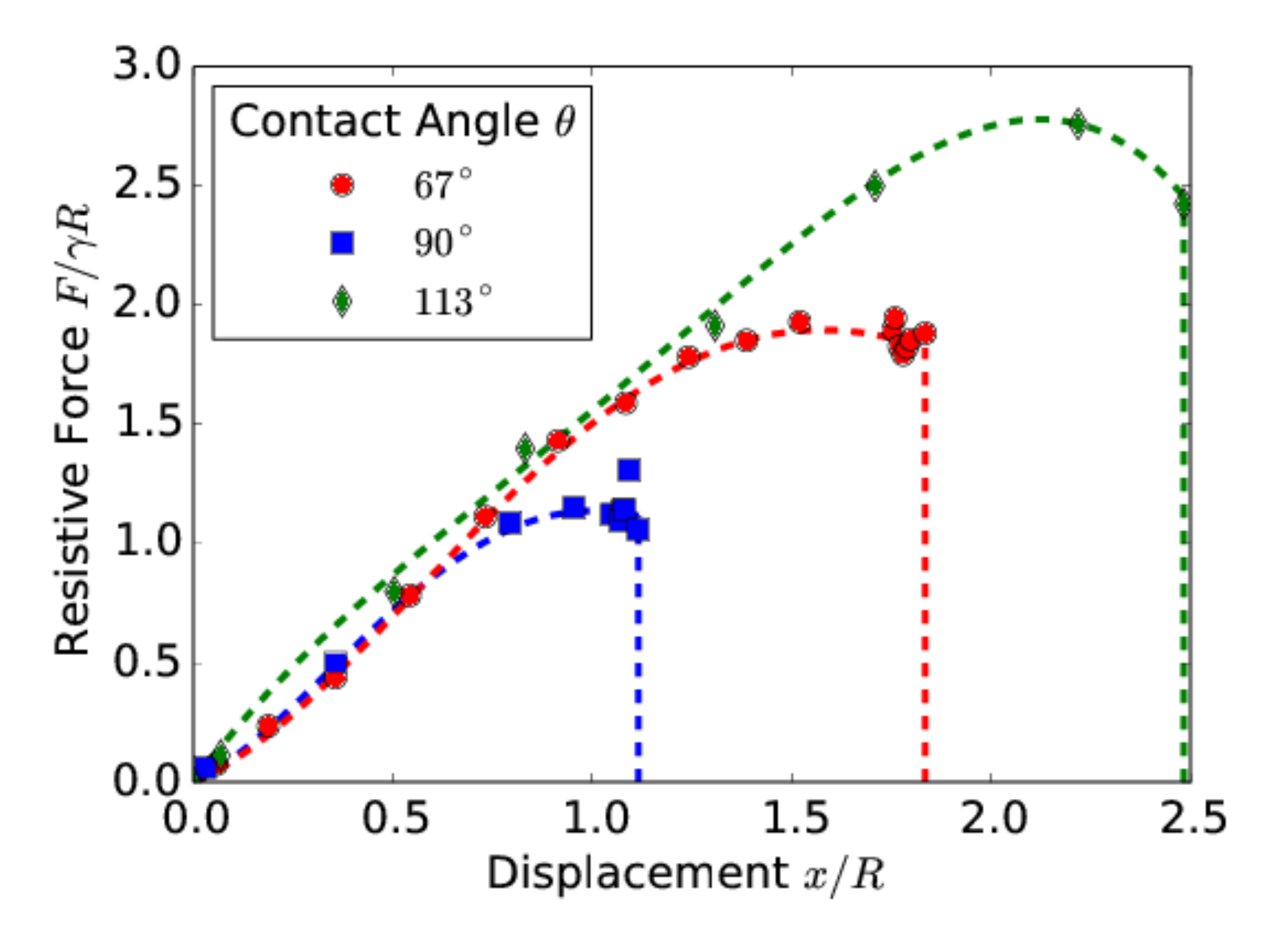}
\caption{\raggedright The normalized resistive force for a spherical particle of radius $R=10$ is approximately linear for small displacements as predicted by de Gennes et al.~\cite{de_gennes_wetting:_1985,joanny_model_1984} and O'Brien.~\cite{obrien_meniscus_1996} The symbols are simulation data and the dashed lines represent fourth-order polynomial fits to that data. The fourth-order fits are integrated in order to obtain the detachment energy, $E$.}
\label{pic:linearforce}
\end{center}
\end{figure}

\begin{figure}
    \centering	
    \begin{subfigure}[t]{\linewidth}	
    \includegraphics[width=\linewidth]{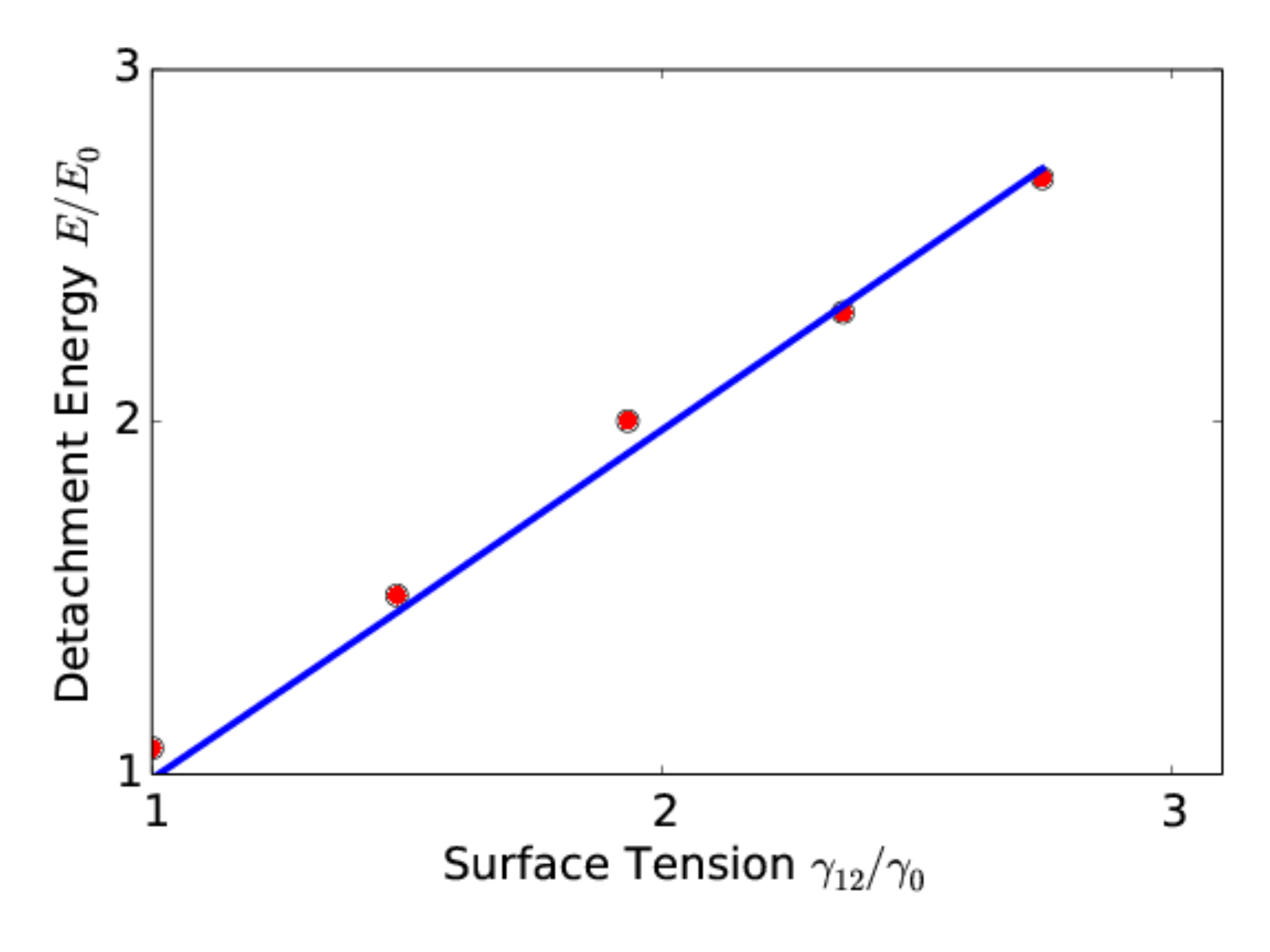}
    \caption{}
    \label{pic:bench1}
    \end{subfigure}
    \begin{subfigure}[t]{\linewidth}
    \includegraphics[width=\linewidth]{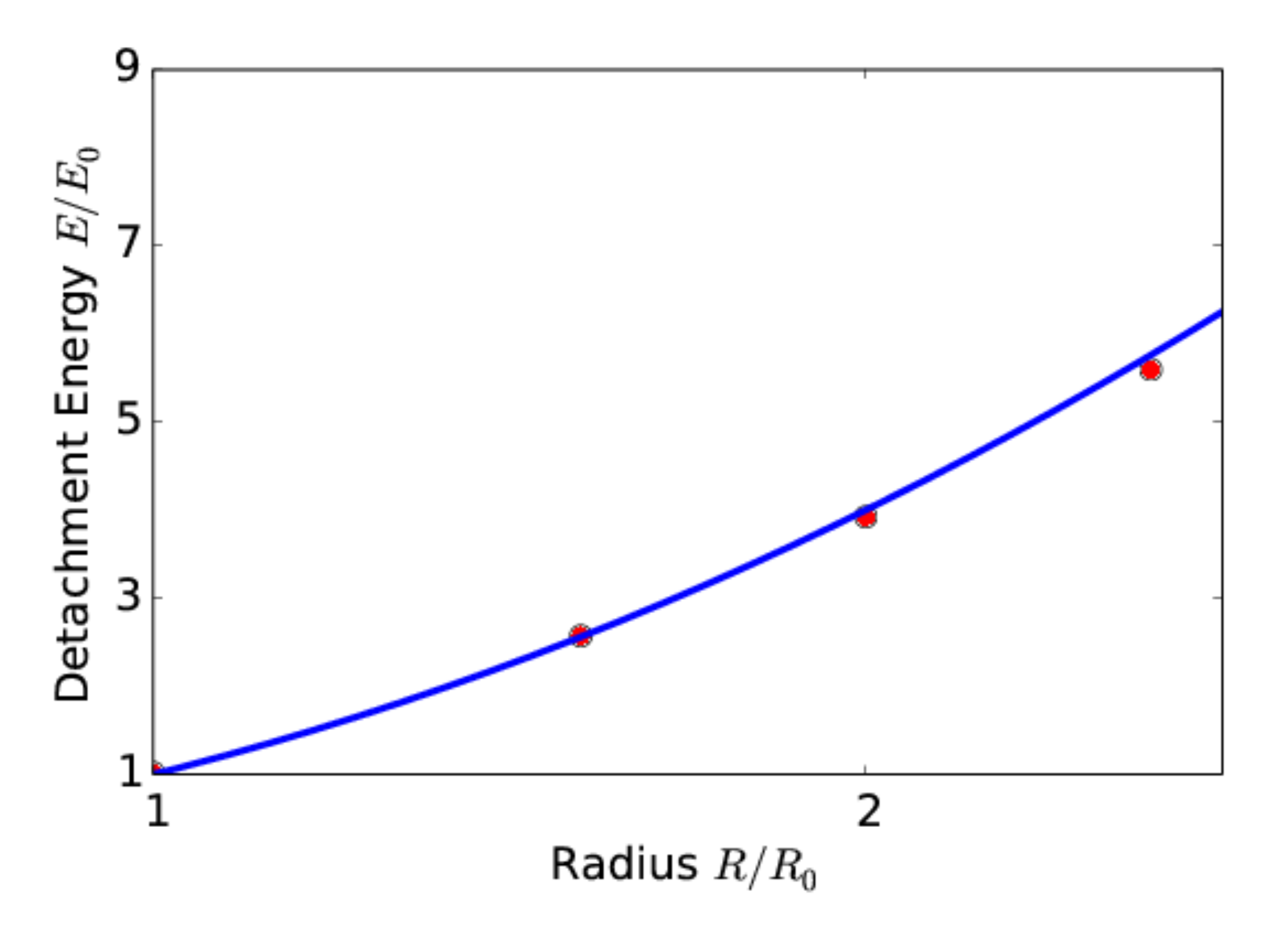}
       \caption{}
    \label{pic:bench2}
    \end{subfigure}
    \caption{\raggedright Variations in the detachment energy as either (a) surface-tension (b) or particle radius is varied. Symbols are data, solid lines are analytical predictions. Error bars are of the order of the symbol size.}
    \label{plot:benchmarks}
\end{figure}
The restoring force provided by the interface to the particle as a function of displacement from equilibrium is shown in Fig.~\ref{pic:linearforce}. The corresponding fourth-order polynomial fits are also shown. The slope of the curves for small displacements are \todo{contact angle} invariant. This agrees with the theoretical calculations of Pitois et al,~\cite{pitois_small_2002} suggesting that the interface stiffness $k$ is a function of the Bond number only. This implies that only the distance at which the particle detaches, and hence the maximum force value, changes with the \todo{contact angle}. However, our data show that there is a significant non-linear regime before the particle detaches, which has a large effect on the detachment energy. 

Figure \ref{pic:bench1} shows the dependence of the detachment energy on the surface-tension for a neutrally wetting particle ($\theta = 90^{\circ}$). $\gamma_0=0.028158$ is the value of the smallest surface-tension we can obtain in our simulations, and $E_0 = \gamma_0 \pi R^2 = 8.846$ is a characteristic energy that corresponds to that surface-tension. We find $\Delta E$ vs $\gamma_{12}$ to be a straight-line with gradient $\pi R^2 (1-|\cos\theta|)^2$ since $\Delta E \propto \gamma_{12}$, as expected.  The values of the surface-tension $\gamma_{12}$ cover the entire accessible range of surface-tensions achievable in the Shan-Chen multicomponent model. We see that the simulation model and the analytical theory agree well for the entire surface-tension parameter range. \\\indent
Figure \ref{pic:bench2} shows a similar plot, this time the variation of the detachment energy as a
function of the particle radius with constant surface tension,
$\gamma_{12}=0.0633$, and \todo{contact angle}, $\theta = 90^{\circ}$. $R_0 = 5$
is the smallest particle radius that produces a well-defined
\todo{contact angle},~\cite{jansen_bijels_2011} and $E_0 =  \gamma \pi R_0^2 = 5.2104$ is a characteristic energy
that corresponds to that particle radius. We find the detachment energy to be a
quadratic function of the radius, $\Delta E \propto R^2$, as expected from the
theory. For a particle of radius $R=5$, the interface thickness equals the
particle radius, yet agreement between our numerical data and the analytical
theory is still excellent, showing that the particle-interface scale
separation does not need to be large, at least for neutrally wetting particles.

\begin{figure}
\begin{center}
\includegraphics[width=8.6cm]{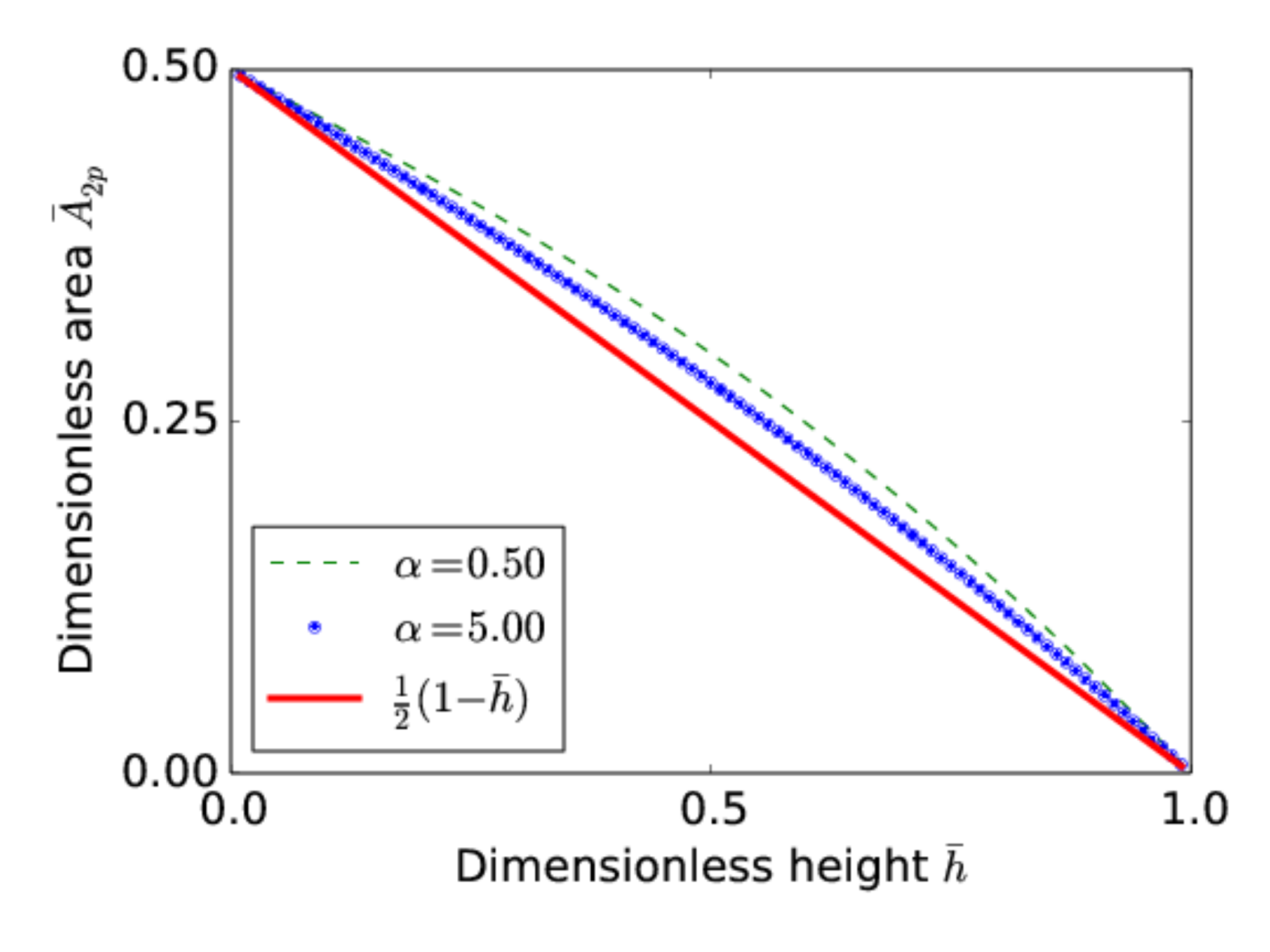}
\caption{\raggedright For $\alpha < 1$, $\bar{A}^{\perp}_{2p}$ (Eq. \ref{ap2perp}) is plotted (dashed line), and for $\alpha > 1$, $\bar{A}^{\parallel}_{2p}$ (Eq. \ref{ap2par}) is plotted (dotted line). The results are compared with the function $A_{2p} = \frac{1}{2}(1-\bar{h})$ (solid line). The functions $\bar{A}^{\perp}_{2p}$ and  $\bar{A}^{\parallel}_{2p}$ are well approximated by the function $\frac{1}{2}(1-\bar{h})$. }
\label{pic:particlearea}
\end{center}
\end{figure}

\subsection{Detachment Energy of Spheroidal Particles}

For spherical particles the contact angle and surface tensions are related according to
\[ \cos \theta = \frac{\gamma_{2p} - \gamma_{1p}}{\gamma_{12}} = \frac{h}{R} := \bar{h}\]

\noindent where $h$ is the height of the particle \todo{centre of mass} above the interface and $R$ the particle radius (Fig.~\ref{pic:contact}). Since measuring the particle-fluid surface tensions experimentally is difficult, we hypothesise that $\frac{\gamma_{2p} - \gamma_{1p}}{\gamma_{12}} \mapsto \bar{h}$ is also a valid substitution for spheroidal particles. This substitution is complicated somewhat for spheroidal particles because there are two potential radii to choose, $R_{\perp}$ or $R_{\|}$. For oblate spheroids in their equilibrium configuration, we choose
\[\frac{\gamma_{2p} - \gamma_{1p}}{\gamma_{12}} \mapsto \frac{h}{R_\|} := \bar{h}\]
and for prolate spheroids in their equilibrium configuration, we take
\[\frac{\gamma_{2p} - \gamma_{1p}}{\gamma_{12}} \mapsto \frac{h}{R_\perp} := \bar{h}.\]
Eq.~\eqref{eq:fperp} and eq.~\eqref{eq:fpar} then become 
\begin{align}
\label{eq:fperp3}
\Delta E^{\perp} & = \frac{\alpha}{4G} \left(1-\bar{h}^2\right) - \bar{h} \bar{A}^{\perp}_{2p} (\bar{h}), \\
\label{eq:fpar3}
\Delta E^{\parallel} & = \frac{1}{4G}\left(1-\bar{h}^2\right) - \bar{h} \bar{A}^{\parallel}_{2p} (\bar{h}).
\end{align}

\noindent In Fig.~\ref{pic:particlearea} we plot the functions $\bar{A}^{\perp}_{2p}(\bar h)$ and $\bar{A}^{\parallel}_{2p}(\bar h)$ as defined in eq.~\eqref{ap2perp} and eq.~\eqref{ap2par}. We see that they are well approximated by the linear function 
\begin{align}
\label{eq:area_approx}
A_{2p} = \frac{1}{2} \left(1-\bar{h}\right).
\end{align}

Incorporating these approximations into our already simplified model in eq.~\eqref{eq:fperp3} and eq.~\eqref{eq:fpar3} yields

\todo{\begin{align}
\label{eq:fperph}
\Delta E^{\perp} & = \frac{\bar{h}^2}{2} \left(1-\frac{\alpha}{2G(\alpha)}\right) - \frac{\bar{h}}{2} + \frac{\alpha}{4G(\alpha)}, \\
\label{eq:fparh}
\Delta E^{\parallel} & = \frac{\bar{h}^2}{2} \left(1-\frac{\alpha}{2G(\alpha)}\right) - \frac{\bar{h}}{2} + \frac{1}{4G(\alpha)}.
\end{align}}
\begin{figure}
\begin{center}
\includegraphics[width=8.6cm]{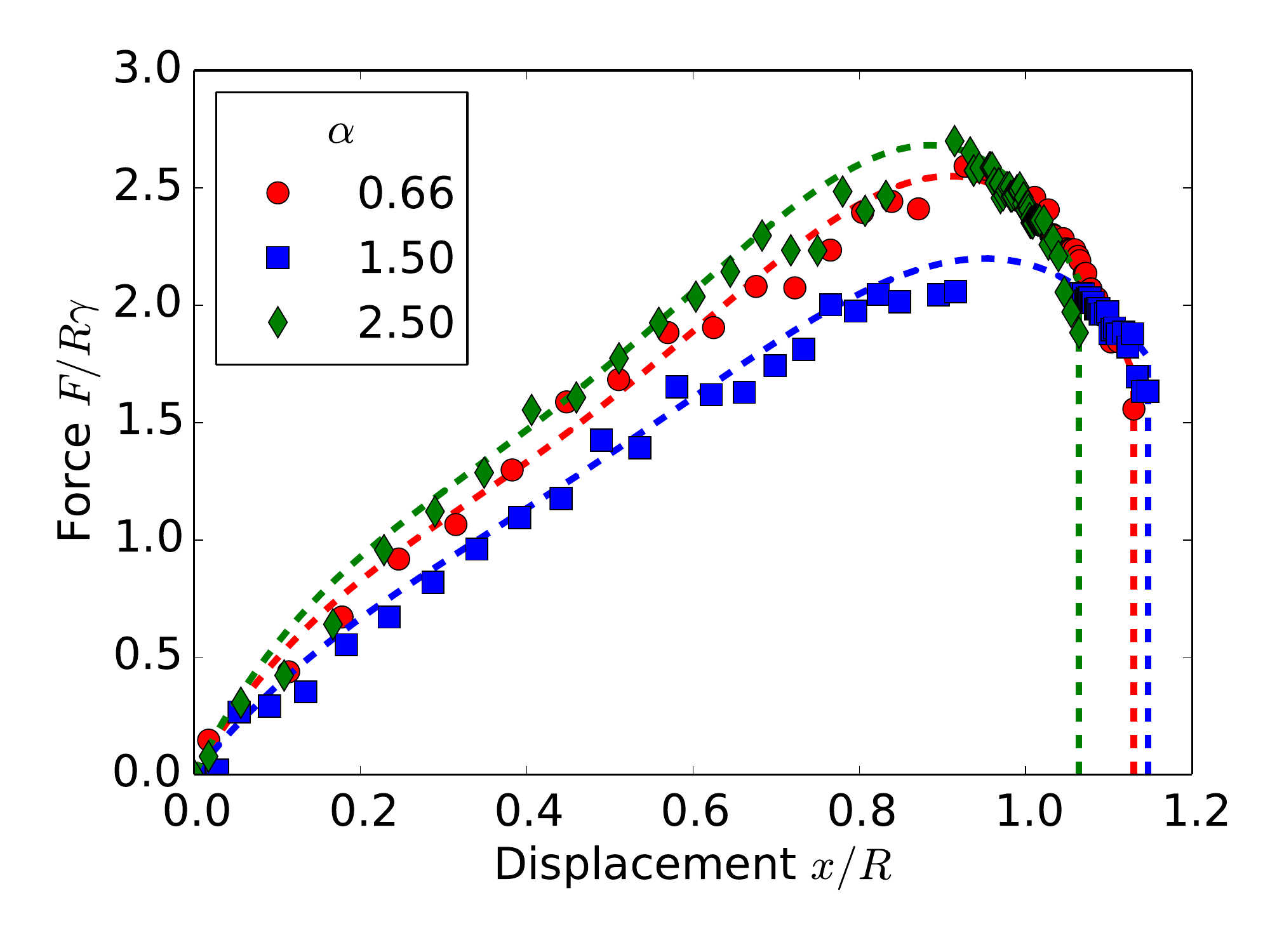}
\caption{\raggedright The interface resistive force on a particle with \todo{contact angle} $\theta=68^{\circ}$ for several aspect ratios $\alpha$. For small displacements, the resistive forces are linear, and for a given displacement, they depend on the aspect ratio. The normalisation factor $R = \sqrt{R_{\perp}R_{\|}}$ and $R = R_{\perp}$ for prolate and oblate spheroidal particles, respectively, is proportional to the area removed by the particle from the interface.}
\label{pic:aninterfaceforce}
\end{center}
\end{figure}
\noindent These approximate expressions are now simple quadratic functions of the dimensionless height $\bar{h}$ and the aspect ratio $\alpha$, and hence eq.~\eqref{eq:fperph} and eq.~\eqref{eq:fparh} represent our simple thermodynamic model describing the detachment energies of prolate and oblate spheroids in their equilibrium positions from interfaces. 

In Fig.~\ref{pic:aninterfaceforce}, we investigate the interface force on anisotropic particles with \todo{contact angle} $\theta=68^{\circ}$ of various aspect ratios. Fig.~\ref{pic:aninterfaceforce} shows that the detachment distance depends weakly on the aspect ratio, but that the resistive force on the particle for a given displacement depends more strongly on the aspect ratio. The resistive interface force is linear until a critical distance whereupon it reaches its peak. Beyond the peak, the force decreases as the particle displacement increases before discontinuously falling to zero --- this is the position where the particle detaches. The slope of the curves depends on the \todo{aspect ratio}, which in the Hookean model suggests a spring constant which depends on the \todo{aspect ratio} in addition to the Bond number. However, our data show that the detachment energy for oblate and prolate spheroidal particles is not as easily modelled using a Hookean approach as it is with spherical particles. The non-linear regime, which suggests the interface has been deformed beyond its elastic limits, has a large contribution to the detachment energy for non-neutrally wetting prolate and oblate spheroidal particles.

We normalise the data using a characteristic particle radius $R = \sqrt{R_{\perp}R_{\|}}$ for prolate and $R = R_{\perp}$ for oblate particles, which are proportional to the area removed from the interface by the particle. We remove outliers related to particle-pinning arising from the staircase approximation of the particles, for the benefit of visualising the data more easily. However, we include these outliers in our fitting function and hence they are taken into account in our numerically calculated detachment energies. 

Comparing Fig.~\ref{pic:linearforce} and Fig.~\ref{pic:aninterfaceforce} we see
that, for spherical particles of varying contact angle, the detachment energy
changes simply because the interface is able to stay attached to the particle
for longer, as the magnitude of the force is equal for small displacements
(Fig.~\ref{pic:linearforce}). In contrast, when the contact angle is constant
but the aspect ratio is varied, the force is different for each aspect ratio
but the particles detach at roughly the same distance for aspect ratios $0.66
\leq \alpha \leq 2.50$. 

\begin{figure}
\begin{center}
\includegraphics[width=8.6cm]{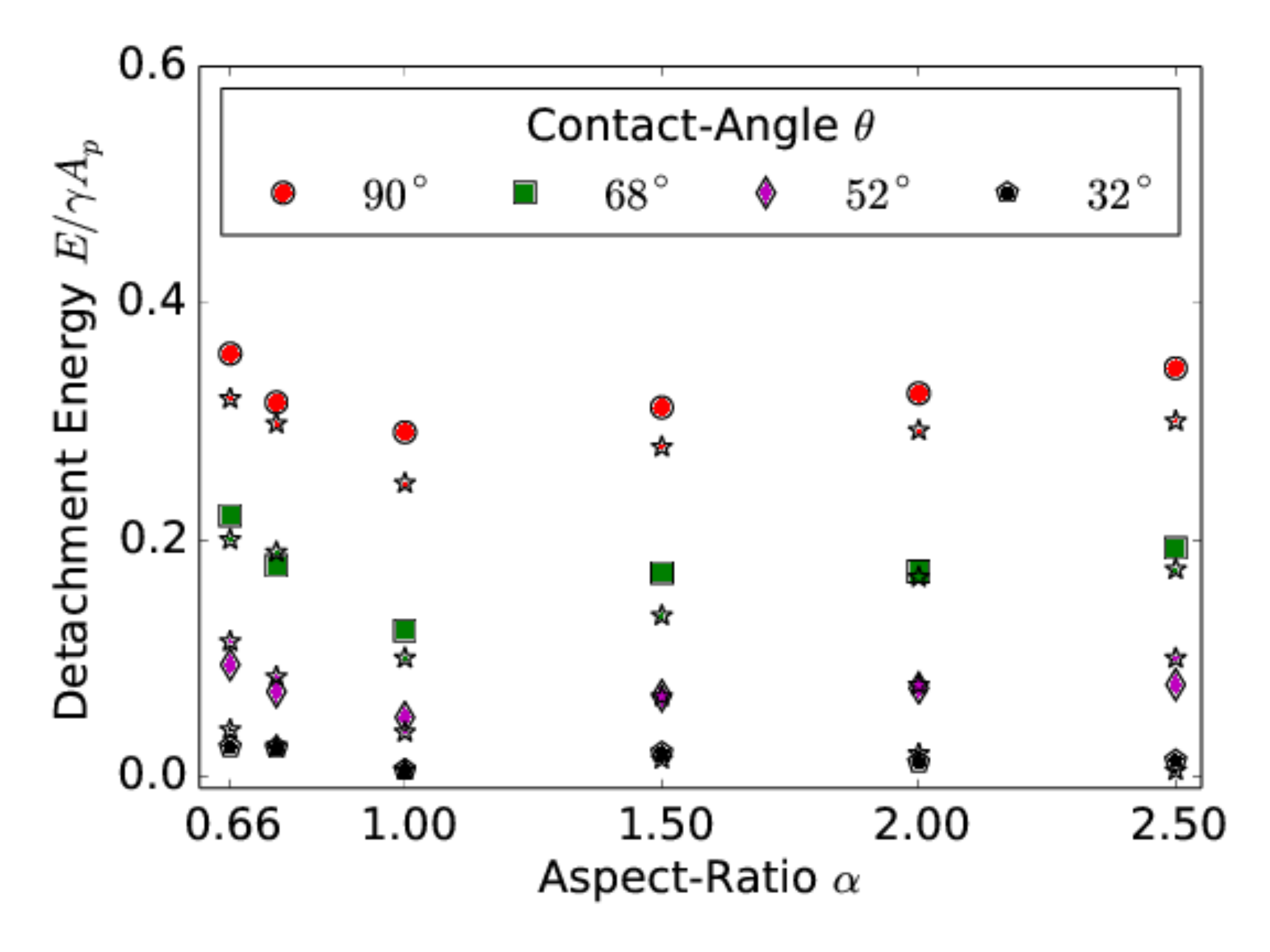}
\caption{\raggedright Dependence of the detachment energy on the aspect ratio, $\alpha$, for several different \todo{contact angles}, $\theta$. Each set of coloured data points represents different wettabilities of the particles. Stars are theoretical calculations from our thermodynamic model in eq.~\eqref{eq:fperph} and eq.~\eqref{eq:fparh} and symbols are numerical data.}
\label{pic:anidetachment}
\end{center}
\end{figure}

In Fig.~\ref{pic:anidetachment} we compare the analytical results from Eq.~\eqref{eq:fperph} and Eq.~\eqref{eq:fparh} with our simulation data. The energy is normalised by the product of the surface-tension, $\gamma$, and the area of the particle, $A_p$. To calculate the theoretical comparison values, we substitute the relevant variables into our model in Eq.~\eqref{eq:fperph} and Eq.~\eqref{eq:fparh}. The dimensionless height is calculated by using the height of the interface far away from the particle. 
We see good agreement between our thermodynamic model and our numerical simulations. For neutrally wetting particles, the measured detachment energies from simulations are larger than those predicted by the thermodynamic model; this is expected since the particle has to deform the interface to overcome its resistive force. The differences are of the order of $10\%$, suggesting that the thermodynamic model, which does not take into account interface deformations, is fairly accurate for the particle \todo{aspect ratio}s we investigated. Similarly, we find good agreement for $\theta=68^{\circ}$ where the numerical data show a higher detachment energy than predicted by the thermodynamic model, as expected. For contact angles $\theta=52^{\circ}$ and $\theta=32^{\circ}$, we still find good qualitative agreement between thermodynamic theory and numerical simulations for both prolate and oblate spheroids. However, for oblate spheroids the numerical detachment energy is less than the analytical predictions, though within errors of the order of the symbol size.   

The success of our thermodynamic model, in particular the validation of the mapping $\frac{\gamma_{2p} - \gamma_{1p}}{\gamma_{12}} \mapsto \bar{h}$, has an important experimental ramification: for prolate spheroidal particles, one needs to measure only the height of the particle \todo{centre of mass} above the interface to determine the contact angle. This should greatly ease contact angle measurements for non-neutrally wetting prolate spheroidal particles, which deform the three-phase contact line, and could be tested experimentally using e.g. a film-calliper method.~\cite{horozov_calliper_2008, arnaudov_measuring_2010}

Further, a recent experiment on the physical ageing of the contact line on colloidal particles at fluid interfaces~\cite{kaz_physical_2012} revealed surprising results, which have only recently been investigated theoretically:~\cite{colosqui_colloidal_2013} particle adsorption involves a sudden breach of the interface followed by relaxation logarithmic in time, showing similarities with ageing in glassy systems. Experiments with oblate and prolate spheroidal particles in which the particle's height above the interface, and hence contact angle, is measured as a function of time may provide insight into this contact line ageing phenomenon.

\section{Conclusions}
\label{chap:conclusions}

We developed a simple thermodynamic model for the detachment energy of prolate and oblate spheroidal particles from fluid-fluid interfaces, which depends on the particle \todo{aspect ratio}, $\alpha = \frac{R_{\|}}{R_{\perp}}$, and the particle dimensionless height, $\bar{h}$, only. $R_{\|}$ and $R_{\perp}$ are the particle radii parallel and perpendicular to the particle's symmetry axis respectively, and $\bar{h} = h/R_{\|}$ and $\bar{h} = h/R_{\perp}$ for oblate and prolate spheroidal particles in their equilibrium position at an interface, respectively. We tested our simple thermodynamic model by detaching spheroidal particles from liquid-liquid interfaces using a Shan-Chen multicomponent lattice Boltzmann model, finding good quantitative and qualitative agreement. 

Our results provide evidence for the validity of our thermodynamic model, supporting our hypothesis that Young's equation $\frac{\gamma_{2p} - \gamma_{1p}}{\gamma_{12}}$ is equal to a suitably defined dimensionless height $\bar{h}$ for spheroidal particles. This result has significant experimental consequences because it should greatly ease \todo{contact angle} measurements for prolate spheroidal particles, which usually deform the three-phase contact-line, making \todo{contact angle} measurements difficult. This prediction may be tested experimentally using e.g. a film-calliper method previously used to measure the \todo{contact angle} of spherical particles.~\cite{horozov_calliper_2008, arnaudov_measuring_2010} Further, using our predicted relation between the height of a prolate particle above the interface and its \todo{contact angle}, surprising experimental results~\cite{kaz_physical_2012} on the physical ageing of contact-lines may be further illuminated by investigating the contact-line ageing of spheroidal particles. \\\indent
There are several natural extensions to the work reported in this paper. Much
research focusses on emulsions and particle-stabilised non-planar liquid-liquid
interfaces. The effect of interface curvature on the detachment energy of
spherical particles has been investigated analytically, though the detachment
energy of anisotropic particles from curved liquid-liquid interfaces has yet to
be extensively investigated.~\cite{aveyard_aspects_2003,komura_adsorption_2006,levine_capillary_1991} It has been suggested that for dimensionless curvatures much less than one, $\frac{R_p}{R_d}
\ll 1$, where $R_p$ and $R_d$ are the particle and droplet radii respectively, Equation (\ref{detachment_energy}) is still valid. Recent experiments~\cite{koos_capillary_2011} investigating the viscosity of suspensions with small volume fractions of immiscible secondary fluid indicate the formation of emulsions droplets with ${R_p} \sim {R_d}$.  Studying the detachment energy as a function of $\frac{R_p}{R_d}$ could provide important insights into the formation of such emulsions.

\begin{acknowledgments}
JH acknowledges financial support from NWO/STW (VIDI grant 10787 of J.
Harting). GBD and PVC thank EPSRC for funding (EPSRC Grant No. EP/I034602/1
“Large Scale Lattice Boltzmann for Biocolloidal Systems”). GBD also thanks
Fujitsu Laboratories Europe for funding an Impact Studentship, and HPC-Europa2
for an award allowing GBD to visit JH at TU/e. TK thanks the University of
Edinburgh for the award of a Chancellor's Fellowship. The authors acknowledge
the use of the UCL \textit{Legion} High Performance Computing Facility
(Legion@UCL), and the UK's national high-performance computing service HECToR,
and associated support services, in the completion of this work. We thank O. Henrich and F. Bresme for useful discussions. 
\end{acknowledgments}

\bibliography{test,tnopapers}

\end{document}